# Robust two-dimensional ice on graphene built from finite-length water molecular chains

Sheng Han[1,§], Jia-Bin Qiao[1,§], Lin-Fang Hou[2,§], Yi-Wen Liu[1], Yu Zhang[1], Zi-Han Guo[1], Long-Jing Yin[1,3], Ya-Ning Ren[1], Wei Ji[2,*] and Lin He[1,*]

[1] Center for Advanced Quantum Studies, Department of Physics, Beijing Normal University, Beijing, 100875, China.
[2] Department of Physics, Renmin University of China, Beijing 100872, China.
[3] Key Laboratory for Micro/Nano Optoelectronic Devices of Ministry of Education & Hunan Provincial Key Laboratory of Low-Dimensional Structural Physics and Devices, School of Physics and Electronics, Hunan University, Changsha 410082, China.
[§]These authors contributed equally to this work.
[*] Correspondence and requests for materials should be addressed to:
Wei Ji (email: wji@ruc.edu.cn) and Lin He (e-mail: helin@bnu.edu.cn)

**Interfacial ice on graphene has attracted much attention because it is a model system to study two-dimensional (2D) ice structures on chemically inert substrates. While water-graphene interaction was usually assumed to be negligible, the structures of the 2D ice are believed to be not appreciably perturbed by the graphene substrate. Here we report atomic-resolved characterizations of an exotic 2D ice structure on graphene built from water molecular chains with finite lengths. Our experiments demonstrated that the water molecular chains are exactly orientated along zigzag directions of the graphene substrate, which evidences an anomalously strong interlayer interaction between the 2D ice and the graphene substrate. Moreover, the length of the water molecular chains closely links to the number of graphene layers, indicating layer-number-dependent water-graphene interfacial interactions. Our work highlights the important role of the 2D ice structures on the water-graphene interfacial interactions.**

Graphene provides an excellent platform to realize precise formation of molecular assemblies and to explore underlying growth mechanism and molecular dynamics because of its homogeneity, flatness and weak screening effect[1-5]. Two dimensional (2D) ice on graphene, as a model system to study adsorption mechanisms and interfacial properties in molecule-solid interface, has attracted much attention because that it is also an idea system in exploring 2D ice structures on chemically inert substrates[6-13]. Unlike those 2D ice structures on metals[14-20], it is believed that the water-graphene interaction is much weaker than the intermolecular interactions of water and, therefore, the structures of the 2D ice are appear to be independent of the adjacent graphene sheets[6,7]. Here, by using scanning tunneling microscope (STM), we report atomic-resolved characterizations of a novel 2D ice structure, which is different from all previously known 2D ice phases, built from water molecular chains with finite lengths. Our experiments demonstrate that these water molecular chains are always oriented along the zigzag directions of graphene and the chain length depends on the number of graphene layers underneath, indicating strong interlayer coupling between the 2D ice and graphene. These results suggest that the water-graphene interactions are much beyond current understandings.

In our experiment, the water-graphene interfaces are obtained in ultrahigh vacuum (UHV) chamber in our STM systems, as schematically shown in Fig. 1a. The water molecules are introduced by wet-etching technique during transferring graphene on target substrates[21-24] and they are inevitably trapped between graphene and the substrates. Our experiments demonstrate that the confined water molecules can escape through boundaries and/or defects of graphene films into the UHV chamber by heating the sample to 300 ~ 400 ℃ for several minutes. Then the base pressure of the chamber increased from about $1.0×10^{-10}$ Torr to $(2 ~ 6)×10^{-9}$ Torr and some of the diffused water molecules thus adsorbed on the graphene surface forming our observed water-graphene interfaces. In our experiment, the pressure in the deposition process of water molecules is much lower than those, on the order of $10^{-5} ~ 10^{-4}$ Torr, reported in previous studies. This could enable the adsorption of water molecules on the graphene surface in a more uniform order to form 2D ice, instead of the formation of amorphous solid water. Once the base pressure of the chamber reenters ~ $1.0×10^{-10}$ Torr, the water covered graphene samples were rapidly cooled down to low temperature (mainly 77 K or 4.2 K). Then the samples were carefully studied by STM at the low temperature. According to our experimental results shown in Fig. 1, the formation of the 2D ice structures is quite robust because that they were observed in all studied graphene systems on different substrates. These systems include as grown graphene on metals Cu (Fig. 1b & 1d) and Ni (Fig. 1e) and high orientated pyrolytic graphite (HOPG, Fig. 1f), as well as transferred graphene on metal Ag (Fig. 1c & 1g) and oxides $SiO_2$ (Fig. 1h) and $SrTiO_3$ (STO) (Fig. 1i)

According to the STM measurements, the apparent thickness of the 2D ice monolayer on graphene is 184.1±9.4 pm (see Supplementary Fig. 1 for more details), which is close to that ~ 160 pm of the monolayer hexagonal water ice on graphite[7]. Our experiment demonstrates that applying a 3V or larger pulse voltage on the STM tip for a sub-second duration can easily break the bonding among the water molecules in the 2D ice and generate artificial void in the 2D ice, which uncovers the graphene lattice underneath (see Supplementary Fig. 2 for more details). By increasing the temperature of the sample to approximate the room temperature, the water molecules of the 2D ice can be gradually removed from the graphene surface and then the graphene lattice fades in (see Supplementary Fig. 3 for more details). All these results explicitly support the formation of novel 2D ice structures on graphene in our experiment.

From the morphologies shown in the Fig. 1, the 2D ice films on different substrates display a striking similarity that they are all comprised of a series of nanoscale water chains. In Fig. 2, we summarize atomic-resolved characterizations of the 2D ice formed by the nanoscale water chains. Figure 2a shows a typical high-resolution STM image of the ice chains on graphene/[Ag(111)/mica]. Apparently, each ice chain (marked by the red dashed frame) is built from buckled triangular water molecules with lattice period $a = (2.50 \pm 0.04)$ Å along the chain direction (see Fig. 2b and 2e), which is quite close to the lattice constant of graphene (~ 2.46 Å). The average chain spacing is measured as $\lambda = (5.0 \pm 0.2)$ Å (see Fig. 2f). Similar lattice period and chain spacing were also observed in all the 2D ice structure on graphene supported by the other substrates under investigation (see Supplementary Fig. 4 for more details). The observed triangular structure in the 2D ice structure is in sharp difference from the conventional puckered hexagonal bilayer ice[20], hexagonal monolayer ice without shared edges[7] observed on HOPG or square monolayer ice[6] confined between two graphene sheets. Especially, the lattice constant of our structure is much smaller than those of previously discovered 2D ice phases[6,7,20,25-36], indicating the formation of a novel triangular 2D ice phase on graphene.

Density functional theory (DFT) calculations were carried out to further explore atomic details of the 2D ice. Three adsorbed structures were considered based on their fully relaxed monolayer geometries (see Supplementary Figs. 5-7 for more details). Among them, the structure shown in Fig. 2b (top view) and 2c (side view) shows the best stability. It is a puckered triangular-ice structure which appears to be highly consistent with our experimental observed structure. A simulated STM image of the puckered structure (Fig. 2d) well captures the chain-like features of the observed water chains (Fig. 2a). In the triangular ice phase, constituent oxygen (O) atoms sit at two different heights, while hydrogen (H) atoms are in a upward-downward staggered arrangement. The lattice period and chain spacing obtained from DFT are $a^* \sim 2.46$ Å and $\lambda^* \sim 5.3$ Å respectively, which are in good agreement with the experimental values of $a = (2.50 \pm 0.04)$ Å and $\lambda = (5.0 \pm 0.2)$ Å.

Given a closer examination of our high-resolution STM images of the adsorbed 2D ice, it was found that the water chains are always oriented along the zigzag direction of the graphene substrate, as shown in Fig. 3a. Therefore, the angle between individual water chains is either 0° or 120° (Fig. 3b, see Supplementary Fig. 8 for more details). Our theoretical calculation also reveals that the adsorbed water chains energetically prefer to align with the zigzag direction rather than the armchair direction (see Supplementary Fig. 9). This orientation preference indicates that the graphene substrate could affect structures of the 2D ice, implying an appreciable interaction between the 2D ice and graphene. In our experiment, we find that the 2D ice structures can grow across the steps of the substrates, as shown in Fig. 3c, indicating that the 2D ice behaves as a flexible 2D material. This is further demonstrated in Fig. 3d and 3e, in which we can clearly observe (sub)nanometer-wavelength rippling in the 2D ice films. Such a feature is similar as the strain-induced nanometer-wavelength ripples in graphene[37-39]. The emergence of the strain-induced ripples in the 2D ice also demonstrates the existence of an appreciable interlayer coupling between the 2D ice and graphene. Previously, it has been demonstrated explicitly that the water-graphene interaction is negligible when the water molecules are in the hexagonal monolayer ice[7] and square monolayer ice phases[6]. Based on these results and our experimental result, it is reasonable to conclude that the water-graphene interactions are dependent of the structures of the 2D ice and are much more complex and beyond current understanding.

Although all the studied 2D ice structures in our experiment have the same lattice period and chain spacing, the lengths of the water chains $L$ are different in these samples, as shown in Fig. 1. The 2D puckered triangular ice is not continuous along the chain direction but "cracked" into piece by piece, featuring an array network organized by nanoscale ice chains in parallel. Figure 3f summarizes the length distribution diagrams of the water chains measured in different samples. It is interesting to find that the lengths of the water chains on mono-, bi- and tri-layer graphene are $L = (13\pm1)a$, however, the lengths become $L = (7\pm1)a$ when the 2D ice is on top of multilayer graphene (with the thickness $n > 3$) or graphite. Such a result suggests that the ice-chain lengths are intimately linked to the layer number of graphene, which implies a layer-number-dependent water-graphene interactions in the studied systems. Although we do not have a complete explanation for this phenomenon, it may be related to the wettability (hydrophilicity or hydrophobicity) of graphene, which is demonstrated to be dependent of the layer number of graphene[40-43].

Besides the monolayer puckered triangular ice on top of graphene, we can also observe bundles of the water chains on the surface of the 2D ice. Figure 4a shows a representative STM image of such a region, where discrete 2, 3, and 4-chain bundles (islands) can be clearly observed on top of the first-layer 2D ice. Our experiment demonstrates that the lattice period, the chain spacing, and the lengths of the water chains of the second-layer 2D ice islands are the same as that of the first layer. By applying a voltage pulse on the STM tip, we generate artificial void in the 2D ice and, simultaneously, generate new second-layer water-chain islands around the void (see Supplementary Fig. 2 for more details). The above results indicate that the growth mode of adsorbed water molecules on graphene turns out to be nucleation and assembling of nanoscale triangular-ice chains and the novel 2D ice structure prefers layer-by-layer growth.

In summary, we report a novel 2D ice, which prefers layer-by-layer growth, built from water molecular chains with finite lengths on graphene. Our experiment demonstrates that the water molecular chains are always along the zigzag directions of graphene and the graphene can introduce strains in the 2D ice, indicating that the interlayer coupling between the 2D ice and graphene is quite strong. Moreover, it is interesting to note that the lengths of the water molecular chains depend on the layer number of graphene. These results suggest that the phenomena of 2D ice on graphene are quite rich and deserve further studies both in experiment and theory, especially with considering the water-graphene interactions.

**Methods**

**Sample preparation**

Our experiments mainly focus on six different substrates: few-layer graphene grown on Ni foils; HOPG; Monolayer graphene transferred on Ag/mica; Monolayer or bilayer graphene grown on Cu foils; Bilayer graphene transferred on $SiO_2$/Si substrate; Trilayer graphene transferred on Nb-doped $SrTiO_3$. The layer number of the graphene is determined by both Raman measurements and STM characterizations.

*Few-layer graphene grown on Ni foils, or p-MLG/Ni*

Few-layer graphene was grown on Ni foils by traditional ambient pressure chemical vapor deposition (APCVD) method with Ar: $H_2$:$CH_4$=10:5:2 at the growth temperature of 1030℃ for 20min.

*HOPG, or p-HOPG*

The highly oriented pyrolytic graphite samples were of ZYA grade (from NT-MDT) and were surface cleaved immediately by the adhesive tape method before experiments without any further processing (e.g., annealing).

*Monolayer/bilayer graphene grown on Cu foils, or p-M(B)LG/Cu*

Monolayer or bilayer graphene was grown on Cu foils by traditional low-pressure chemical vapor deposition (LPCVD) method with Ar: $H_2$:$CH_4$=20:10:1 at the growth temperature of 1035℃ for 30min.

*Monolayer graphene transferred on Ag/mica substrates, or t-MLG/Ag*

Monolayer graphene grown on Cu foil was transferred on Ag/mica substrates *via* well-established PMMA-assisted wet etching techniques. In short, PMMA is evenly coated onto the graphene surface on the copper substrate by a Spin Coater, and then heated and dried. After treatment, the sample is put into saturated ammonium persulfate until the copper was completely corroded, leaving only free-floating graphene with PMMA. The samples are removed to dilute hydrochloric acid for washing for about 2 hours to ensure that ammonium persulfate is completely reacted, followed by multiple washing with deionized water. Graphene suspending in deionized water is then collected on target substrate and dried on a heating plate. Transferred samples are annealed at around 500℃ for 12 hours in a low-pressure (~ $10^{-3}$ Torr) tube furnace to remove PMMA and other organic residues and then quickly transferred into UHV chamber for further annealing.

*Bilayer graphene transferred on SiO$_2$/Si substrate, or t-BLG/SiO$_2$*

Monolayer graphene grown on copper foil was transferred on the SiO$_2$/Si substrate *via* well-established PMMA-assisted wet etching techniques for twice, and the impurities were removed by tube-furnace annealing.

*Trilayer graphene transferred on Nb-doped SrTiO$_3$, or t-TLG/STO*

Monolayer graphene grown on copper foil was transferred to the Nb-doped SrTiO$_3$

substrate *via* well-established PMMA-assisted wet etching techniques for three times, and the impurities were removed by tube-furnace annealing.

**UHV annealing and STM experiment**

Ultrahigh vacuum scanning tunneling microscope (STM) systems (USM-1300, USM-1400 and USM-1500) from UNISOKO was utilized for UHV annealing and STM measurements. The specimens were quickly transferred into the UHV chamber of STM system after tube-furnace annealing process. The UHV annealing was performed at chamber pressure within $(2 \sim 6) \times 10^{-9}$ Torr and $T$ within $300 \sim 400$ ℃ for $\sim 10$ mins. When the base pressure of the chamber reenters $\sim 1.0 \times 10^{-10}$ Torr and the temperature of sample holder returns to room temperature after deposition, the specimens were rapidly cooled down to liquid nitrogen temperature ($\sim 77$ K) and lower liquid helium temperature ($\sim 4.2$ K) for the assembling of adsorbed water molecules into wetting layer and eventually ice crystals. Then low-$T$ STM characterization were conducted on the specimens when the temperature and pressure of the probing system become stable.

The STM tips were obtained by chemical etching from a wire of Pt(80%) Ir(20%) alloys. STM measurement was performed in the ultrahigh vacuum chamber ($\sim 10^{-11}$ Torr) with constant-current scanning mode mainly at $\sim 77$ K or $\sim 4.2$ K. Lateral dimensions observed in the STM images were calibrated using a standard graphene lattice as well as a Si(111)-(7×7)lattice and Ag(111) surface.

**DFT calculations**

Density functional theory (DFT) calculations were performed using the generalized gradient approximation and the projector augmented wave method[44,45] as implemented in the Vienna *ab initio* simulation package (VASP)[46,47]. The uniform Monkhorst-Pack k mesh of 2×10×1 was adopted for integration over the Brillouin zone (BZ). A plane-wave cutoff energy of 400 eV was used for all calculations. All atoms were allowed to relax until the residual force on each atom was less than 0.02 eV/Å and the bottom layer C substrate was fixed in the DFT calculations. A vacuum layer of 20 Å in thickness was employed. Van der Waals interactions were considered at the vdW-DF level[48,49] with the optB86b[44] functional for the exchange potential (optB86b-vdW), which was proved to be accurate in describing structural related properties of layered materials[51-56]. In simulation, the graphitic substrate was simulated by bilayer graphene with the bottom layer fixed.

**Acknowledgements**

This work was supported by the National Natural Science Foundation of China (Grant Nos. 11974050, 11674029). L.H. also acknowledges support from the National Program for Support of Top-notch Young Professionals, support from "the Fundamental Research Funds for the Central Universities", and support from "Chang Jiang Scholars Program".

**Author contributions**

S.H., J.B.Q., Y.W.L., Y.Z., Z.H.G., L.J.Y. and Y.N.R. synthesized the samples and performed the STM experiments. L.F.H. carried out the DFT calculations. S.H., J.B.Q., and L.F.H. analyzed the data. L.H. conceived and provided advice on the experiment, analysis, and theoretical calculation. L.H., S.H., J.B.Q. and L.F.H. wrote the manuscript with the input from all authors.

**Competing interests**

The authors declare no competing financial interests.

**Figures**

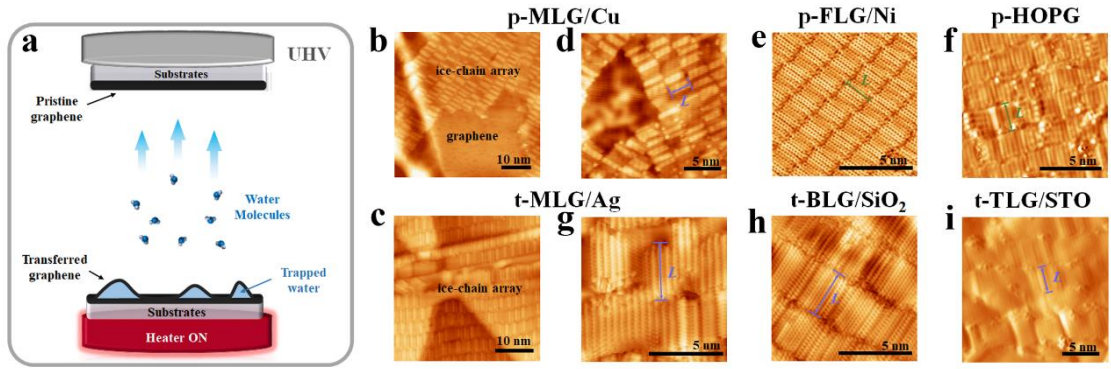

**Figure 1 | Assembling of ice-chain array on various graphene-based substrates. a**, Cartoon of the UHV annealing serving as water deposition technique. Water molecules escaping from the trapping of graphene films during the process are adsorbed and arranged on surfaces of both transferred and pristine graphene sheets. **b**, Large-scale STM image ($V_b$ = 500 mV, $I$ = 100 pA) exhibiting 2D ice-chain array overlayers on monolayer graphene CVD-grown on Cu foils. **c**, Large-scale STM image ($V_b$ = 600 mV, $I$ = 200 pA) displaying the ice-chain arrays grown on transferred monolayer graphene on Ag(111)/mica. **d-i**, STM close-ups of 2D ice-chain array adlayers on monolayer graphene grown on Cu foils (600 mV, 100 pA), multilayer graphene grown on Ni foils (544 mV, 100 pA), HOPG (700 mV, 200 pA), monolayer graphene on Ag(111)/mica (600 mV, 200 pA), bilayer graphene on SiO$_2$/Si (600 mV, 200 pA) and trilayer graphene on Nb-doped SrTiO$_3$ (1.7 V, 100 pA), respectively. The lengths of the water chains, as marked by the green arrows in **e** and **f** and the blue ones in **d**, **g-i**, are different and are summarized in Fig. 3f. All STM images were taken at $T$ ~ 4.2 K.

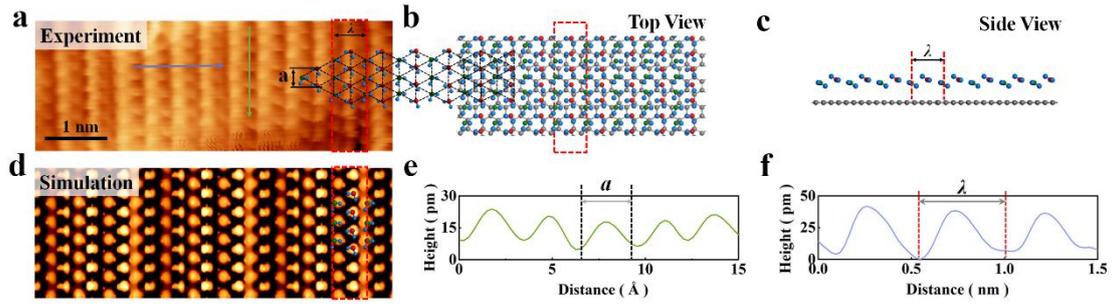

**Figure 2 | Triangular ice chains on graphene. a**, Atomic-resolution STM topography ($T \sim 4.2$ K, $V_b = 600$ mV, $I = 200$ pA) of the ice-chain array on transferred monolayer graphene on Ag(111)/mica, exhibiting the constituent triangular water molecules with lattice period $a$ and chain spacing $\lambda$. **b**, Top view of the simulated configuration of ice chain on graphene lattice along the zigzag direction. Grey spheres correspond to C atoms, red ones to upper O atoms, green ones to lower O atoms and small blue ones to H atoms. Overlay of the configuration on STM image in **a** exhibits excellent consistency. **c**, Side view of the simulated ice-chain configuration. **d**, DFT simulation of the STM topography based on the configuration in **b**. Height profiles along the green (**e**) and blue (**f**) arrows marked in **a**, indicating the lattice period $a$ and chain spacing $\lambda$, respectively. The red dashed frames in **a**, **b** and **d** indicate the single ice chain.

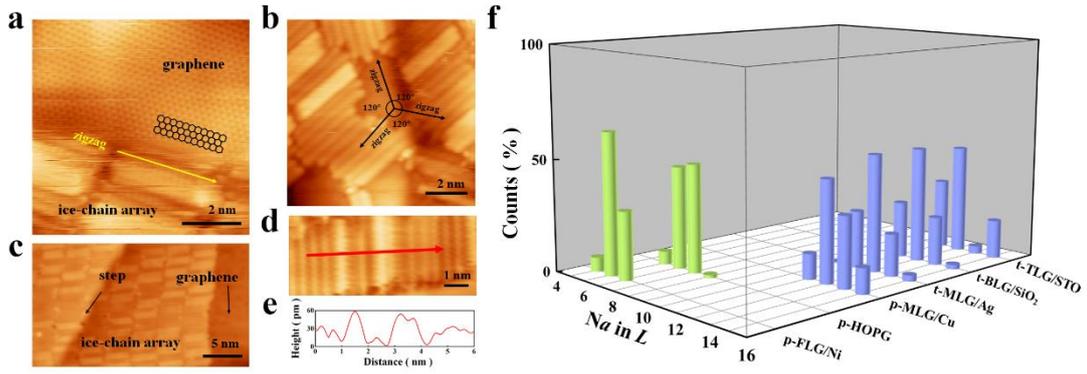

**Figure 3 | STM characterization of the nanoscale ice chains on graphene. a**, Atomically resolved STM image ($T \sim 4.2$ K, $V_b = 800$ mV, $I = 100$ pA) displaying the ice-chain arrays grown on transferred monolayer graphene on Ag(111)/mica along the zigzag direction of graphene lattice. **b**, Large-scale STM image ($T \sim 4.2$ K, $V_b = 800$ mV, $I = 100$ pA) showing that the ice-chain structures in the 2D ice are always along the zigzag direction of graphene. Therefore, the relative angle between the ice chains is either 0° or 120°. **c**, Large-scale STM image ($T \sim 4.2$ K, $V_b = 500$ mV, $I = 100$ pA) showing the arrays arranged in parallel and across the graphene/Cu step. **d,** Typical high-resolution STM image ($T \sim 4.2$ K, $V_b = 600$ mV, $I = 200$ pA) of nanoscale ice chains arranged in parallel on t-MLG/Ag. **e**, Height profile along the red arrow marked in **d**. We can clearly observe (sub)nanometer-wavelength rippling in the 2D ice. **f**, Three-dimensional histogram of the length of chain $L$ in unit of lattice period $a$ extracted from hundreds of ice chains on six different graphene-based substrates, revealing the layer-number-dependence of the length of nanoscale ice chains on different substrates.

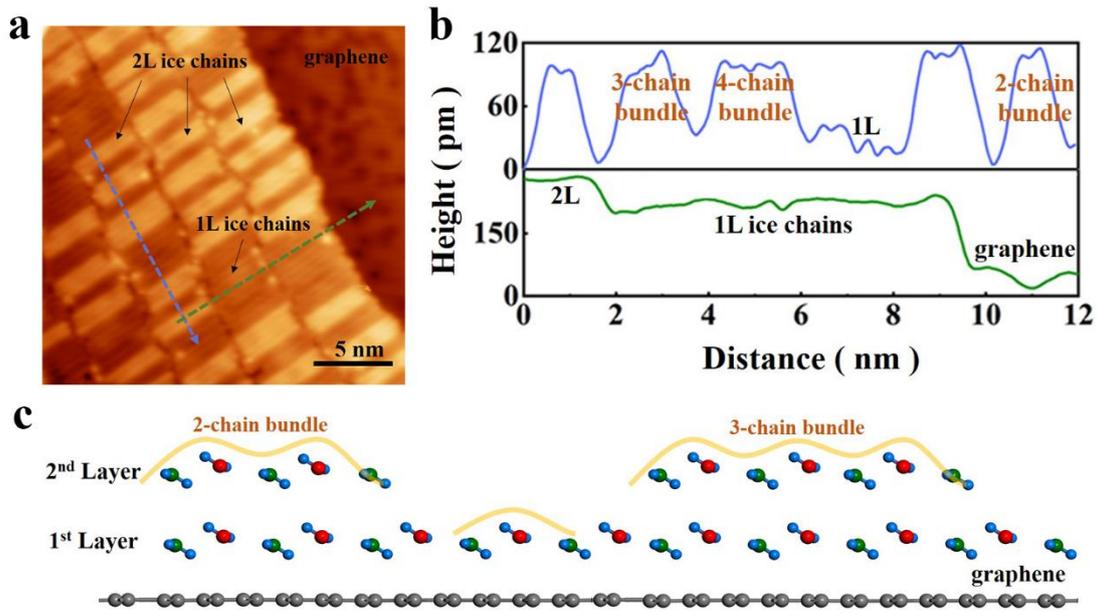

**Figure 4 | Characterization of bundle-like bilayer ice chains on graphene. a**, STM image of ice-chain array overlayers on monolayer graphene grown on Cu foils ($T \sim 4.2$ K, $V_b = 500$mV, $I = 100$ pA), showing the area of mixed mono- and bi-layer ice-chain adlayer on graphene surface. **b**, Height profiles along the blue and green lines marked in **a**, presenting various bilayer ice-chain bundles on the monolayer ice-chain array. **c**, Side view of the schematic of bilayer 2, 3-chain bundles and monolayer ice-chain array grown on graphene lattice.